\documentstyle[psfig]{mn}

\def\rmi{\max[R_0,R_{\rm if}(t)]}
\def\rma{\max[R_{\rm M},R_{\rm if}(t)]}
\def\ltsima{$\; \buildrel < \over \sim \;$}
\def\lsim{\lower.5ex\hbox{\ltsima}}
\def\gtsima{$\; \buildrel > \over \sim \;$}
\def\gsim{\lower.5ex\hbox{\gtsima}}

\begin{document}

\title[Time dependent opacities]
{Time dependent photoionization opacities in dense Gamma--Ray Burst
environments}

\author[Lazzati, Perna \& Ghisellini]
{Davide Lazzati$^1$, Rosalba Perna$^{2,3}$ 
\& Gabriele Ghisellini$^4$\\
$^1$ Institute of Astronomy, University of Cambridge, Madingley Road, 
Cambridge CB3 0HA, England\\
$^2$ Harvard Society of Fellows, 78 Mt. Auburn Street, Cambridge, MA 02138\\ 
$^3$ Harvard--Smithsonian Center for Astrophysics,
60 Garden Street, Cambridge, MA 02138\\
$^4$ Osservatorio Astronomico di Brera, via Bianchi 46, 23807 Merate (LC), 
Italy\\
{\tt e-mail: lazzati@ast.cam.ac.uk}}

\maketitle

\begin{abstract}
The recent detection of a transient absorption feature in the X--ray
prompt emission of GRB~990705 showed the importance of such
observations in the understanding of gamma--ray bursts and their
progenitors. We investigate the time dependence of photoionization
edges during the prompt emission of bursts in different environments.
We show that their variability can be used to infer the density and
geometry of the surrounding medium, giving important clues to unveil
the nature of the burst progenitor.
\end{abstract}


\section{Introduction}

The nature of the progenitors of gamma--ray bursts (GRBs) is still
mysterious, since the fireball producing the gamma--ray emission does
not carry any information about the progenitor generating it (see,
e.g., Piran 1999).  Recent observations (e.g., Kulkarni et al. 1999)
suggest the association of GRBs with the final stages of evolution of
massive stars (Woosley 1993, Paczy\'nski 1998; MacFadyen \& Woosley
1999), but many crucial points still remain controversial.  The
possible evidence of rebrightening shown by the optical afterglows of
several bursts $\sim30$ days after the burst explosion (Bloom et
al. 1999; Reichart 1999; Galama et al. 2000) suggests a simultaneous
explosion of the burst with a Type Ic supernova.  On the other hand,
the detection of iron absorption and emission features in the X--ray
emission of the bursts and afterglows (Amati et al. 2000; Piro et
al. 2000) can be explained with a two--step explosion, in which the
burst onset follows a supernova explosion by several months (Lazzati
et al. 1999; Vietri et al. 2001; Lazzati et al. 2001).

These different scenarios can be distinguished through the analysis of
the radial distribution of the interstellar medium surrounding the
explosion site (Ghisellini et al. 1999).  If the burst explodes
simultaneously with the supernova, the photons propagate through the
pre--explosion stellar wind, while in the two--step scenario a high
density metal enriched supernova remnant is expected to surround the
burst explosion site (Lazzati et al. 1999).

In this paper we study the variability properties of photoionization
edges in the first several tens of seconds from the burst onset, for
the different radial distributions of the density profile
characterizing these two scenarios. Temporal variations of the
opacities are indeed expected as a result of the gradual
photoionization of the medium by the X-ray prompt emission of the
burst.  A similar variability effect on absorption lines had been
discussed by Perna \& Loeb (1998) and, in the X--rays, by B\"ottcher
et al. (1999). In these papers, however, dependences on the radial
profile of the density had not been considered.

The paper is organized as follows: in \S2, we present some analytic
approximations for the time-dependent opacities which hold for the
optically thin case, discussed in \S3.  In \S 4, we present the
results of the general (numerical) solution to the problem, and
compare it with the analytic approximations derived in \S2.  In \S5,
we discuss possible effects of the fireball expansion in the absorbing
medium.  Finally, our results are summarized in \S6.

\section{Analytic theory}

Consider a distribution $n(r)$ of absorbers, characterized by a
photoionization cross--section $\sigma$. In the case of an impulsive
illumination by a strong ionizing flux the density $n(r)$ becomes time
dependent, and so does the absorption opacity:
\begin{equation}
\tau(t) = \sigma \, \int_0^\infty n(r,t)\,dr
\label{eq:taut}
\end{equation}
To solve Eq.~\ref{eq:taut}, we must determine the time--dependent
density $n(r,t)$. We use a particular time reference: at each radius
$r$ from the photon source, the time is set to be zero when the first
ionizing photon passes by.  By saying that the ionization time at
radius $\tilde r$ is $\tilde t$, we then mean that $\tilde t$ is the
time interval between the first ionizing photon crossing the radius
$\tilde r$ and the complete ionization of all absorbing ions located
at the same radius.  The advantage of this time reference is that the
travel time of photons is null, and the time $t$ is also the time
measured by an observer at infinity.  The same conceptual scheme
applies in \S 5 for the radius of the fireball.

Consider now a medium in which recombination is not efficient and let
us neglect the change of the resonant frequency as an atom is
progressively stripped off its electrons from the ground state to
complete ionization. The first condition is easily fulfilled in GRBs,
given their high ionization flux during the prompt phase (exceptions
may however be possible, see Lazzati et al. 2001).  In this case the
density $n(r,t)$ remains constant until all electrons are stripped.
The time of complete stripping depends on the distance $r$ of an atom
from the photon source. We assume that, at a given radius $\tilde r$,
$n(\tilde r,t)$ changes abruptly in time:
\begin{equation}
n(\tilde r,t) = n(\tilde r)\, \chi_{[0,\tilde Z\,t_{\rm ion}(\tilde r)]}(t)
\label{eq:nstep}
\end{equation}
where $ \chi_{[a,b]}(x) = 1$ for $a \le x \le b$ and 0 elsewhere,
$t_{\rm ion}$ is the ionization time of a single electron and $\tilde
Z$ is the ``efficient number" of electrons of the relative element.
$\tilde Z$ is in general smaller than the atomic number $Z$ of the
element, since other processes can contribute to ionization.  These
include collisional ionization, the Auger effect and the ionization of
the external electrons by the UV flux.  The actual value of $\tilde Z$
can be obtained through numerical simulations and is not universal,
being dependent on the ionizing UV continuum and the density of the
absorbing medium (see \S~4).  The ionization time of a single electron
is given by:
\begin{equation}
t_{\rm ion} = \left[ \int_{\nu_0}^\infty {{F(\nu)}\over{h\,\nu}}\,
\sigma(\nu)\,d\nu\right]^{-1}
\label{eq:tion1}
\end{equation}
where $F(\nu)$ is the flux of ionizing radiation at a given radius,
taking into account the effect of absorption from material at lower
radii.  Eq.~\ref{eq:tion1} modifies as:
\begin{eqnarray}
t_{\rm ion}(r) &=& {{4\pi\,r^2}\over{
\int_{\nu_0}^\infty {{L(\nu)}\over{h\,\nu}}\,e^{-\tau(\nu)}\,
\sigma(\nu)\,d\nu}} = \nonumber \\
&=& {{4\pi\,r^2}\over{
\int_{\nu_0}^\infty {{L(\nu)}\over{h\,\nu}}\,e^{-\sigma(\nu)
\int_0^r n(\rho,t)\,d\rho}\,
\sigma(\nu)\,d\nu}}
\label{eq:tion2}
\end{eqnarray}
where $L(\nu)$ is the luminosity of the ionizing continuum and $\nu_0$
is the threshold photoionization frequency.  Note that the ionization
time, needed to compute the density of absorbers, depends on the
density itself in a non--linear way.  For this reason
Eq.~\ref{eq:tion2} does not have a general analytical solution.

\begin{figure}
\psfig{file=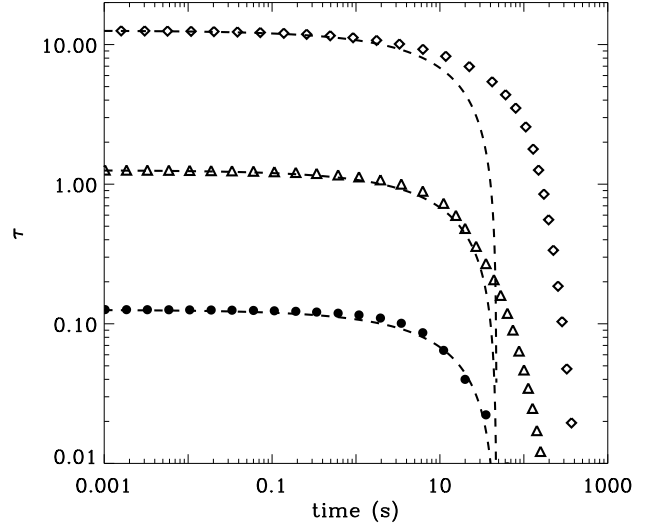,width=0.47\textwidth}
\caption{{Opacity vs. time for a uniform cloud of absorbers. 
Initially opaque, intermediate and thin clouds are considered, from
top to bottom (see text for the parameter set).  The dashed lines
correspond to the analytic approximations of Eq.~\ref{eq:tauuni},
while the symbols are obtained with the numerical simulations
described in \S 4.  As expected, the analytic approximation holds in
the optically thin regime, but fails to reproduce accurately the drop
of the opacity in the optically thick cases (top line and diamonds).
}
\label{fig1}}
\end{figure}

\section{Optically thin media}

If the absorbing medium is optically thin, the flux at a given radius
does not depend (at least as a first approximation) on the absorption
taking place at smaller radii.  In this case Eq.~\ref{eq:tion2}
simplifies to:
\begin{equation}
t_{\rm ion}(r) = {{4\pi\,r^2}\over{
\int_{\nu_0}^\infty {{L(\nu)}\over{h\,\nu}}\,\sigma(\nu)\,d\nu}}
\label{eq:tion3}
\end{equation}
By approximating the shape of a photoionization edge as $\sigma(\nu) =
\sigma_0 \, (\nu/\nu_0)^{-3}$, and considering a power--law spectrum
for the ionizing flux [$L(\nu) = L_0\,(\nu/\nu_0)^{-\alpha}$],
Eq.~\ref{eq:tion3} becomes:
\begin{equation}
t_{\rm ion}(r) = (3+\alpha)\,h\,{{4\pi\,r^2}\over{L_0\,\sigma_0}}
\label{eq:tion4}
\end{equation}
Combining Eq.~\ref{eq:tion4} with Eq.~\ref{eq:taut} and
Eq.~\ref{eq:nstep} we obtain:
\begin{equation}
\tau(\nu,t) = \sigma_0\,\left({\nu\over\nu_0}\right)^{-3}\,
\int_{R_{\rm if}(t)}^\infty n(r)\,dr
\end{equation}
where ${R_{\rm if}(t)}$, the radius of the ionization front at time
$t$, is obtained by inverting Eq.~\ref{eq:tion4}:
\begin{equation}
R_{\rm if}(t)= \left( {{L_0\,\sigma_0\,t}\over{4\pi\,h\,
\tilde Z\,(3+\alpha)}}
\right)^{1/2}
\label{eq:rif}
\end{equation}
Assuming an absorbing medium whose density decreases as a power law
with distance, [$n(r)=n_0\,(r/R_0)^{-\beta}$], we obtain
\begin{equation}
\tau(\nu,t) = \sigma_0\,\left({\nu\over\nu_0}\right)^{-3}\,n_0\,R_0^\beta\,
\int_{\rmi}^{\rma} r^{-\beta}\,dr
\label{eq:tau2}
\end{equation}
where $R_{\rm M}$ and $R_0$ are the maximum and minimum radii of the
density profile.  This avoids unphysical divergences at very large and
very small radii.  With this assumptions Eq.~\ref{eq:tau2} becomes:
\begin{eqnarray}
\tau(\nu,t) \!\!\! &=& \!\!\! \sigma_0\,n_0\,R_0^\beta\,
\left({\nu\over\nu_0}\right)^{-3} \times \nonumber \\
\!\!\! &\times& \!\!\!\!\!\!  \left\{
\begin{array}{ll}
\ln\left\{{{\rma}\over{\rmi}}\right\} & \beta = 1 \\
{{{\rma}^{1-\beta} - {\rmi}^{1-\beta}}\over{1-\beta}}
& \beta \ne 1 \end{array} \right.
\label{eq:tau3}
\end{eqnarray}
Two particularly interesting cases are a uniform distribution up to a
given radius $R_{\rm M}$ and a rapidly decreasing ($\beta > 1$)
distribution extending to infinity.

\subsection{Uniform density}

In the case of a uniform density of absorbers ($\beta = 0$),
Eq.~\ref{eq:tau3} becomes particularly simple:
\begin{equation}
\tau(\nu,t) = \sigma_0 n_0 \left({\nu\over\nu_0}\right)^{-3}
\left\{ \begin{array}{ll}
\!\!\!\! R_{\rm M} - R_0 & R_{\rm if} < R_0 \\
\!\!\!\! R_{\rm M} - \sqrt{A\,t} & R_0 < R_{\rm if} < R_{\rm M} \\
\!\!\!\! 0 & R_{\rm if} > R_{\rm M} \end{array} \right.
\label{eq:tauuni}
\end{equation}
where $A$ is given by:
\begin{equation}
A = {{L_0\,\sigma_0}\over{4\pi\,h\,\tilde Z\,(3+\alpha)}}
\label{eq:a}
\end{equation}
Note that the case of uniform density can correspond not only to the
case of a burst surrounded by a standard interstellar medium, but
would also be a reasonable approximation for the case of a young
supernova remnant (see Lazzati et al. 1999).

The opacity is constant at the beginning, then decays becoming rapidly
negligible.  Examples of this behavior are shown in Fig.~\ref{fig1}
(dashed lines) for three different initial conditions: an opaque
($\tau(0) \sim 13$) cloud, an intermediate ($\tau(0) \sim 1.3$) cloud
and a thin ($\tau(0) \sim 0.13$) cloud.  We have considered the
photoionization edge of iron, with a resonance cross section $\sigma_0
\sim 2\times 10^{-20}$~cm$^2$ and a threshold frequency $h\,\nu =
9.28$~keV.  The parameterization adopted for the ionizing flux and the
ambient medium is: $L_0 = 1.75\times10^{31}$~erg~s$^{-1}$~Hz$^{-1}$,
$\alpha = 0$ (similarly to the spectrum of GRB 990705, Amati et
al. 2000), $R_0 = 10^{13}$~cm, $R_{\rm M} = 1$~pc and $n_0 = 200$,
$20$, $2$~cm$^{-3}$, respectively.  These conditions are quite
extreme, since they correspond to a total mass of the absorbing iron
of $\sim 1000$, $100$ and $10\,M_\odot$, respectively.  On the other
hand, a $\tau \sim 1$ feature lasting for $\sim 10$ seconds has indeed
been detected (Amati et al. 2000), and very large iron masses are
required, in some scenarios, to explain this feature (see Lazzati et
al. 2001 and references therein for a more complete discussion).

We have considered in the simulation a constant ionizing luminosity.
This is not generally true for GRBs, where fluctuations on timescales
of fractions of seconds are usually observed.  This would cause small
scale fluctuations, with the same timescale, on the time evolution of
the opacities, but the general trend is unaffected.  Moreover, the
burst may turn off before the complete ionization of the surrounding
medium.  In this case the results of this paper would hold only for
times smaller than the turn off time of the GRB.

By comparison with the numerical simulations presented in \S 4 (where
the approximations of \S2 are released), the ``efficient number" of
electrons of iron in these conditions turns out to be $\tilde Z_{\rm
Fe} \sim 7$.  The dashed lines have been computed adopting this value.

In Fig.~\ref{fig2}, we show the time of complete ionization of the
iron ions as a function of the distance from the source of the
ionizing photons.  In the analytic approximation of Eq.~\ref{eq:rif},
this time does not depend on the initial opacity and density
distribution of the medium.

\subsection{Decaying density profiles}

Let us consider now the case of a density profile decreasing with
distance from the burst source.  This can correspond, for example, to
a pre--explosion stellar wind if the bursts are associated with the
death of massive stars.  In this case we would have $\beta = 2$
(Chevalier \& Li, 1999).

If the density of the absorbing medium decreases with distance
sufficiently rapidly ($\beta > 1$), the maximum radius can be set to
infinity and Eq.~\ref{eq:tau3} becomes:
\begin{equation}
\tau(\nu,t) = 
{{\sigma_0\,n_0\,R_0^\beta\,\left({\nu\over\nu_0}\right)^{-3}}
\over{\beta-1}} \left\{ \begin{array}{ll}
R_0^{1-\beta} & R_{\rm if} < R_0 \\
(A\,t)^{{{1-\beta}\over{2}}} & R_{\rm if} > R_0 
\end{array}\right.
\label{eq:tauwind}
\end{equation}
where $A$ is given in Eq.~\ref{eq:a}.  After a short initial constant
phase, the opacity decreases in time as a power--law whose slope
depends only on the index $\beta$ of the density profile.  The time
evolution of the opacity in a wind environment is shown in
Fig.~\ref{fig3}.  We used $R_0 = 10^{13}$~cm and initial densities
$n_0=10^9$, $10^8$ and $10^7 {\rm cm}^{-3}$ (corresponding to $\tau(0)
\sim 200$, 20 and 2).  The parameters describing the ionization flux
are the same used for Fig.~\ref{fig1}.  In Fig.~\ref{fig4}, we show
the time evolution of the radius of complete ionization, for the same
parameters used for Fig.~\ref{fig3}.

\section{Numerical results}

\begin{figure}
\psfig{file=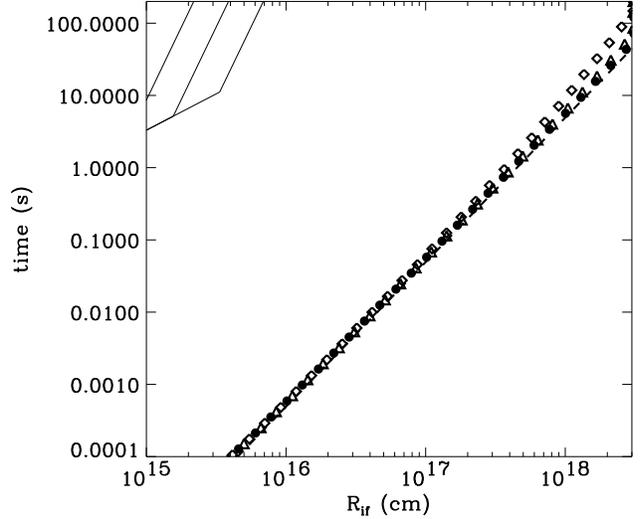,width=0.47\textwidth}
\caption{{Time of complete ionization vs. radius for the same
geometrical setup of Fig.~\ref{fig1}.  The dashed line shows the
analytic approximation of Eq.~\ref{eq:rif}.  The symbols refer to the
same cases of Fig.~\ref{fig1}.  Since in the simulations the
approximation of the step function (Eq.~\ref{eq:nstep}) is not used,
the symbols are relative to the time at which the fraction of iron
completely ionized is a half of the total.  The thin solid lines in
the upper left corner of the figure show the time at which the
fireball crosses the radius $R_{\rm if}$ (see \S 5).  Since this time
is always much larger than the time of ionization, the hydrodynamics
of the fireball does not influence the evolution of the opacity.}
\label{fig2}}
\end{figure}

To solve the problem in its full generality, we remove the
approximations made in \S~2.  We start the simulation at $t=~0$, and
let the burst photons propagate.  In propagating from a point at
position $r$ to another point at position $r+\Delta r$, the ionizing
flux is reduced according to:
\begin{equation}
F_\nu(r+\Delta r,t+\Delta t) = F_\nu(r,t)\exp [-\Delta \tau_\nu(r,t)]
\frac{r^2}{(r+\Delta r)^2}\;.
\label{eq:flux}
\end{equation}
The optical depth due to photoabsorption within the distance $\Delta
r$ is then given by:
\begin{equation}
\Delta \tau_\nu(r,t)= \Delta r \sum_{j}n_j(r,t)\sigma_j(\nu)\;.
\label{eq:tau}
\end{equation}
The photoionization cross sections are taken from Reilman \& Manson
(1979).  The ionic concentrations are determined by solving the system
of equations:
\begin{eqnarray}
\frac{dn_j(r,t)}{dt}&=&q_{j-2}n_{j-2}+ q_{j-1}n_{j-1}
+ c_{j-1}n_{j-1} n_e \nonumber \\ 
&-&(q_j+c_j n_e +\alpha_j n_e)n_j 
+\alpha_{j+1}n_{j+1} n_e\;.
\label{eq:dndt}
\end{eqnarray}
The $q_j$ and $c_j$ are respectively the photoionization and
collisional ionization coefficients of ion $j$, while $\alpha_j$ is
the recombination coefficient. Note that $q_{j-2}$ refers to inner
shell photoionization followed by Auger ionization.  The collisional
ionization rates are calculated according to Younger (1981). We
compute the terms due to photoionization by integrating
$L_\nu\sigma_\nu$ numerically.  The recombination rates are given by
the sum of the radiative and dielectronic recombination rates.  The
code uses routines developed by Raymond (1979; see also Perna, Raymond
\& Loeb 2000 for a similar application and further details).  After
updating the ionization fractions at each time step $\Delta t$, the
optical depth $\tau_\nu(t)=\int dr \Delta\tau_\nu(r,t)$ is recomputed.

\begin{figure}
\psfig{file=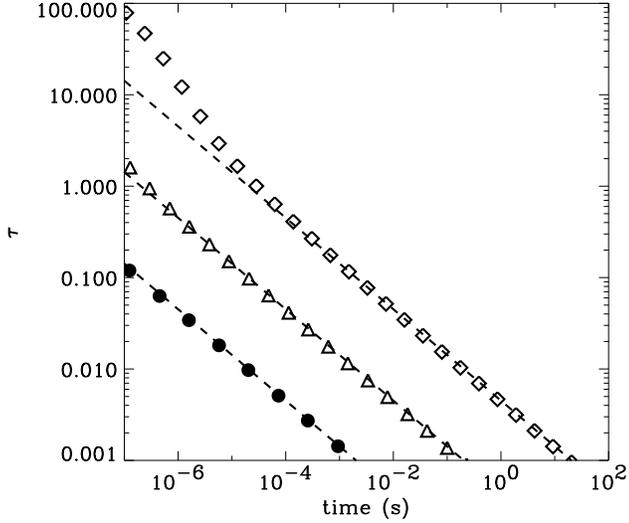,width=0.47\textwidth}
\caption{{Opacity vs. time for a wind environment. 
Initially opaque, intermediate and thin winds are considered, from top
to bottom (see text for the parameter set). The dashed lines
correspond to the analytic approximations of Eq.~\ref{eq:tauwind},
while the symbols are obtained with the numerical simulations
described in \S 4. As expected, the analytic approximation holds for
$\tau < 1$.}
\label{fig3}}
\end{figure}

\begin{figure}
\psfig{file=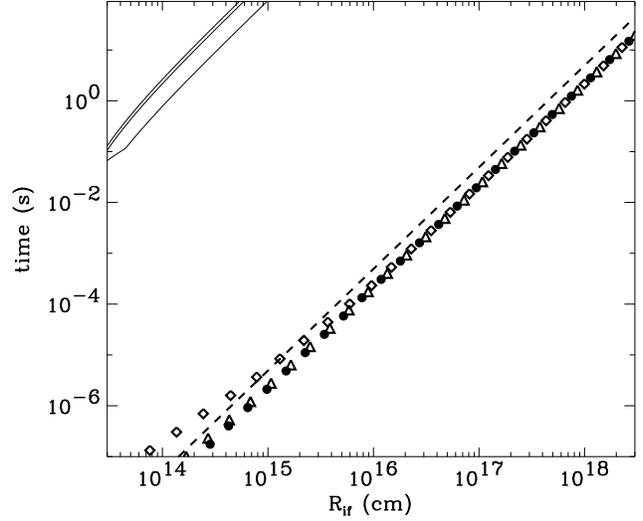,width=0.47\textwidth}
\caption{{Same as Fig.~\ref{fig2} but for the wind environment
of Fig.~\ref{fig3}.}
\label{fig4}}
\end{figure}

A comparison between the numerical and analytical results is shown in
Fig.~\ref{fig1}, \ref{fig2}, \ref{fig3} and~\ref{fig4}. In all these
figures, the analytical approximation is drawn as a dashed line, while
the numerical results are overlaid as symbols (circles, diamonds and
triangles).  The introduction of recombination and collisional
ionization does not change significantly the results with respect to
the analytic case, where only photoionization is considered.  As
expected, corrections must be applied in the optically thick regime.

In Fig.~\ref{fig1}, the opacity is dominated by material at large
radii.  The analytic approximation is valid when the initial opacity
is small or intermediate ($\tau(0) \lsim 1$), but underestimates the
time of decline for an initially opaque cloud. For both the initially
thin and thick cases, the analytic approximation declines faster than
the numerical result. This is due to the approximation of
Eq.~\ref{eq:nstep}, which severely underestimates the absorption at a
given radius for $t > t_{\rm ion}(r)$.

Fig.~\ref{fig2} shows the time of complete ionization for the same
parameter set of Fig.~\ref{fig1}. In the numerical simulations the
time of complete ionization is not a well defined quantity, since at
each radius and at any time there is always the possibility (albeit
very small) to have ions with bounded electrons. To properly compare
with the time defined in Eq.~\ref{eq:tion4}, which is plotted as a
dashed line, the symbols (circles, diamonds and triangles) show the
time at which about 50 \% of the iron is completely stripped
(FeXXVII).  The figure shows that the analytic approximation
reproduces accurately the numeric results, with small deviations at
large radii for the initially opaque case.

In the case of a wind environment (Fig.~\ref{fig3}), the opacity is
dominated by material at small radii.  The analytic approximation
describes accurately the temporal decay of the opacity in the
optically thin and intermediate cases ($\tau(0) \lsim 1$).  For an
initially optically thick medium, the numerical solution deviates at
small times, but collapses onto the analytic approximation when the
medium becomes optically thin.  Again, $\tilde Z_{\rm Fe} \sim 7$ is
adequate.  The analytic time of complete ionization for a wind
environment is compared with the results of numerical simulations in
Fig.~\ref{fig4}.  Again, the analytic approximation is adequately
accurate at all times but for the initially optically thick case, in
which deviations are observed.

\section{Fireball dynamics}

For all our calculations we have assumed that the ionization flux
comes from a point source.  This is not strictly true, since in the
fireball model the prompt radiation is produced by the expanding
shell.  To check whether our assumption is reasonable we must
demonstrate that the fireball radius is smaller than the ionization
radius at all times.

The radius of the relativistic fireball expanding in an inhomogeneous
external medium is given by (see, e.g., Meszaros, Rees \& Wijers,
1998):
\begin{equation}
R_{\rm fb}(t) = R_{\rm IS}+ c \Gamma_0^2 \left \{ 
\begin{array}{ll}
t & t \le t_{\rm ES} \\
t_{\rm ES}^{{{3-\beta}\over{4-\beta}}}\,t^{{{1}\over{4-\beta}}} &
t > t_{\rm ES}
\end{array} \right.\;,
\end{equation}
where $\Gamma_0$ is the asymptotic Lorentz factor of the fireball and
$R_{\rm IS}\sim 10^{13}$~cm is the internal shock radius, i.e. the
radius at which the first ionizing photons are produced.  The time
$t_{\rm ES}$ is the onset time of external shocks, where the fireball
expansion starts to be slowed down by the interaction with the
external medium. This is given by (Meszaros, Rees \& Wijers 1998):
\begin{equation}
t_{\rm ES} = \left\{\left[{{(3-\beta)\,E}\over
{4\pi\,\Gamma_0^2\,m_p\,c^2\,n_{\{H,0\}}}}+R_0^3\right]\,R_0^{-\beta}
\right\}^{1\over{3-\beta}}\, {1\over{\Gamma_0^2\,c}},
\end{equation}
where $E$ is the total fireball energy, $m_p$ the proton mass, and a
particle distribution $n_H (R) = n_{\{H,0\}} (r/R_0)^{-\beta}$ has
been assumed.  Note that the distribution of particles in the external
medium follows the distribution of the absorbers given above.  The
normalization are instead different.  For solar iron abundance we have
$n_0 = 4.68\times10^{-5} n_{\{H,0\}}$ (Anders \& Grevesse 1989).

In Fig.~\ref{fig2} and~\ref{fig4}, the time at which the fireball
crosses the radius $R_{\rm if}$ is plotted with thin solid lines.  We
assumed a fireball with $E=10^{52}$~erg and $\Gamma_0=100$, and an
iron abundance ten times solar. It can be seen that the fireball
crossing time is always much larger than the time of ionization.  We
conclude that, for a reasonable set of parameters, the assumption of
decoupling between the ionization and the hydrodynamics is valid.

\section{Conclusions} 

We have calculated the time dependence of the absorption opacities due
to the material in the surroundings of gamma--ray bursts.  We have
been able to qualitatively reproduce the numerical results with
analitic expressions using some approximations appropriate for the
optically thin regime.

In the optically thick regime, corrections must be applied.  In the
case of a uniform medium, the analytic expressions fail to reproduce
exactly the lifetime of the absorption feature (i.e. the timescale of
constant optical depth), while in the wind case the analytic
expressions are in good agreement with the numerical results at all
but the very early times.

An important issue is the observability of these features.  On one
hand, we require the density of the surrounding medium to be high, in
order to have the highest possible opacity.  On the other hand
(Lazzati et al. 2001), the Thomson scattering optical depth, $\tau_T$,
of the absorbing material must be less than one, in order to maintain
the flickering behavior of the $\gamma$--ray lightcurve. For a given
Thomson opacity, the iron opacity is given by $\tau_{\rm Fe} \sim 1.5
\tau_T A_{\rm Fe}$, where $A_{\rm Fe}$ is the iron abundance in solar
units. Iron enriched media are hence necessary to observe a deep iron
feature in a $\tau_T \lsim 0.1$ cloud.  Enriched media around GRBs
have indeed been observed (Amati et al. 2000; Lazzati et al. 2001).

For the wind environment, the absorption feature has an extremely
short lifetime, not detectable with the present and near future
detectors and instruments.  It then appears that a stellar wind, even
if strong and appropriate to a very massive progenitor, is not able to
imprint a detectable absorption feature in the X--ray spectrum of
bursts.  The reason is the immediate ionization of the absorbing
material.  However, there is still the possibility that the wind of
the progenitor, especially in the last phases, is unsteady and
intermittent, therefore populating the burst environment with a
density profile different from $R^{-2}$ and corresponding to more mass
at larger radii than the case we assumed.  If this case can be
described by a larger value of $R_0$, then the lifetime of the
absorption feature should scale as $R_0^2$ and detectable features may
appear if $R_0\sim10^{15}$ cm and/or with a weaker ionization flux,
i.e. in X--ray poor bursts.  If, beyond $R_0$, the density profile
retains a power law profile, then also the absorption optical depth
decreases in time as a power law, and this discriminates the wind case
from the uniform density scenario, where the fall off of the
absorption feature should be more abrupt.

The required time resolved X--ray spectroscopic observations are
beyond the capabilities of present instruments and satellites, which
can provide only a time integrated measurement of the opacity on
timescales of $5\div10$ seconds (cfr. Amati et al. 2000).  The Swift
satellite, however, will be able to slew on target and to start the
observations with the X--ray telescope in a few seconds (in
particularly favorable cases), and will be sensitive enough to measure
time--resolved opacities (Perna et al., in preparation).

\end{document}